\newcommand{\be}{\begin{equation}}
\newcommand{\ee}{\end{equation}}
\newcommand{\bea}{\begin{eqnarray}}
\newcommand{\eea}{\end{eqnarray}}

\newcommand{\lab}[1]{\label{#1}}

\newcommand{\aleq}{\mathrel{\rlap{\lower4pt\hbox{\hskip1pt$\sim$}}
                   \raise1pt\hbox{$<$}}}
\newcommand{\ageq}{\mathrel{\rlap{\lower4pt\hbox{\hskip1pt$\sim$}}
                   \raise1pt\hbox{$>$}}}
\newcommand{\GeV}{\,{\rm GeV}}

\documentclass[prd,nofootinbib]{revtex4}

\usepackage{indentfirst}
\usepackage{latexsym}
\usepackage{graphicx}

\begin{document}
\title{Natural Inflation, Planck Scale Physics\\
 and Oscillating Primordial Spectrum}
\author{Xiulian Wang$^a$, Bo Feng$^a$, Mingzhe Li$^a$, Xue-Lei Chen$^{b,c}$, Xinmin Zhang$^a$}
\affiliation{$^{a}$ Institute of High Energy Physics, Chinese
Academy of Science, P.O. Box 918-4, Beijing 100039, P.R. China}
\affiliation{$^{b}$ National Astronomical Observatories, Chinese
Academy of Sciences, 20A Datun Road, Beijing 100012, P. R. China}
\affiliation{$^{c}$ Kavli Institute for Theoretical Physics UCSB,
CA 93106, USA }

\begin{abstract}
In the ``natural inflation'' model, the inflaton potential is
periodic. We show that Planck scale physics may induce corrections
to the inflaton potential, which is also periodic with a greater
frequency. Such high frequency corrections produce oscillating
features in the primordial fluctuation power spectrum, which are
not entirely excluded by the current observations and may be
detectable in high precision data of cosmic microwave background
(CMB) anisotropy and large scale structure (LSS) observations.
\end{abstract}

\maketitle

\section{Introduction}
In the past decade, inflation theory has successfully passed
several non-trivial tests. In particular, recent cosmic microwave
background (CMB) observations show that the spatial geometry of
the observable Universe is very close to flat \cite{de
Bernardis00,Lange00,Hanany00,Balbi00,Jaffe00}, just as inflation
theory predicts. Inflation theory also offers an elegant way of
generating the primordial fluctuations which seed the formation of
galaxies and large scale structures (LSS)(see e.g.,
Ref.~\cite{Lyth-Riotto99} and references therein). In particular,
slow-roll inflation models predict that the perturbations are
adiabatic, gaussian, and nearly scale-invariant (i.e with a power
index $n_s \simeq 1$). It has long been known that these
predictions are also in broad agreement with the observed
properties of large scale structures and CMB anisotropy, although
at present the data is still not very restrictive \cite{Wang01}:
$n_s=0.91^{+0.15}_{-0.07}$.

This general success of inflation theory brings up the hope of
extracting even more detailed information of the inflaton
potential\cite{Lyth-Riotto99,Lidsey97} from high quality
observational data. There are two complementary approaches to this
problem. In the first approach, one tries a model-independent
(subject to some standard assumptions) reconstruction of the
primordial power spectrum, then the inflaton potential from the
data \cite{Hannestad01,MSY01,Wang-Mathews02}. Alternatively, one
can look for specific features in the power spectrum and study
their observational consequences. In particular, there have been
many investigations on inflationary models with broken scale
invariance\cite{KLS85,Starobinsky92,ARS97,LPS98,Chung00,WK00,L00}.
Such features have been invoked to explain the tentatively
observed feature at $k \sim 0.05
Mpc^{-1}$\cite{GSZ00,HHV01,BGSS00,GH01}, or even to solve the
small scale problem of the CDM model \cite{KL00}.\footnote{For
other solutions to this problem, see Refs.~\cite{Spergel
PRL}-\cite{Bode APJ} and for a recent review on this issue see
Ref.\cite{Tasitsiomi}}

In the present work, we consider a new type of feature, periodic
in the primordial power spectrum. This type of feature is
interesting from both a theoretical and phenomenological point of
view. Theoretically, if discovered, it gives a strong hint on the
nature of the inflaton field. Phenomenologically, it might change
the position, shape or even the number of acoustic peaks in the
CMB power spectrum.

This paper is organized as follows: in Sec. II, we describe our
model, and show how this type of feature could arise from Planck
scale physics by constructing a toy model. Our toy model, which is
based on the ``natural inflation'' model, is by no means the only
possibility, but in this context it is particularly easy to see
how this might happen. In Sec. III we derive the power spectrum in
this model, and then consider how it would affect CMB and large
scale structure in Sec. IV.  The final section, Sec.V is on the
summary and discussions of our results.

\section{The Model}

In addition to phenomenological success, a compelling inflation
model should also be based on plausible particle physics. To be in
agreement with observations, the inflaton potential must be very
flat. Since the radiative corrections to a scalar field mass are
quadratically divergent, some physical mechanism, e.g., a symmetry
is required to maintain the flatness of the potential, unless we
want to accept ad hoc fine tuning.

For the natural inflation model \cite{FFO90},
several possible physical mechanisms \cite{ABFFO93}
are available for producing a inflaton potential of the form
\begin{equation}
\label{eq:potentiala}
V(\phi)=\Lambda^4\left[1 \pm \cos(\phi/f)\right].
\end{equation}
Customarily, the positive sign is taken, with identical results
for the negative sign. If $\Lambda \sim 10^{16} \GeV$ and
$f\sim10^{19}\GeV$, which are the grand unification scale and
Planck scale respectively, a successful inflation model could be
obtained, with the correct quantum fluctuation amplitudes
\cite{MT01}.

In this paper we extend the natural inflation model above and
include a term in the potential in Eq.(\ref{eq:potentiala}) which
is also periodic but with greater frequency:
\begin{equation}
\label{eq:potentialb}
 V_{planck} (\phi) = \delta \Lambda^{4} \cos(N \phi
/f + \theta ) .
\end{equation}
We will argue below that this term may come from
the Planck scale physics.

In natural inflation, one introduces a pseudo Nambu-Goldstone
boson (PNGB) as the inflaton. Specifically let's consider an
anomalous global $U(1)$ symmetry which is spontaneously broken at
scale  $f$ {\it via} a non-vanishing value of
 a complex scalar field $
\Phi \sim \frac{1}{\sqrt{2}} f e^{i\phi /f}$ with $\phi$ the
Nambu-Goldstone boson. At lower energy scale $\Lambda$ like Axion
a potential for the Goldstone boson $\phi$ is generated from the
non-perturbative effects and its mass is order of $m_{\phi} \sim
\frac{\Lambda^2}{f}$.

Recently there are a lot of interests in studying the effects of
the Planck scale physics on the primordial spectrum and CMB
anisotropy \cite{laoB}. In the original paper of this subject,
Martin and Brandenberger considered a modified dispersion relation
and showed explicitly a sizable effect. In this paper we take an
effective lagrange approach to new physics and argue that Planck
scale physics can correct the inflaton potential, generate
oscillating and scale-dependent power spectrum which may be
detectable in the future.

The Planck scale physics which can be parameterized by higher
dimensional operators makes two types of contributions to the
effective lagrangian of the inflaton $\phi$. Since $\phi$ is the
phase of the scalar field $\Phi$, if the Planck scale physics
preserves the $U(1)$ symmetry the Goldstone theorem requires that
there must be derivatives involved in the operators with $\phi$.
The operators with the lowest dimension are like $\frac{\partial^2
\phi
\partial ^2 \phi}{M_{pl}^2}$. The authors of Ref.\cite{kaloper} have studied
the effects of this type of operators on CMB and shown that the
corrections to CMB is order of $H^2_{inflation}/ M^2_{planck} \sim
10^{-11}$. And Ref.\cite{shiu} pointed out that higher order terms
like $\frac{{( \partial_\mu \phi
\partial^\mu \phi )}^2}{M_{pl}^4}$ would be possible to enhance the
effects. In this paper we study another type of the Planck scale
physics effects where the higher dimensional operators involve no
derivatives and modify the inflaton potential. This requires an
explicit violation of the global symmetry U(1). In fact, as argued
in Refs.\cite{kamion} the quantum gravitational interactions break
the global symmetries and one expects the existence of a type of
higher dimension operators like
\begin{equation}
{O_{planck}}(\Phi,\Phi^{\dag})=g_{MN}\frac{(\Phi^{\dag}\Phi)^M
\Phi^N }{M_{Pl}^{2M+N-4}} +h.c.,
\end{equation}
which gives rise to a correction to the inflaton potential of $\phi$:
\begin{equation}
V_{planck}(\phi) = \delta \Lambda^{4} \cos(N\phi /f + \theta ),
\end{equation}
where
\begin{equation}
\delta  = 2^{-M-N/2+1} \left| g_{MN} \right|
\frac{{M_{Pl}^4}}{\Lambda^4} \left( \frac{f}{M_{Pl}} \right) ^{2M+N}
\end{equation}
with $\theta$ the phase of the parameter $g_{MN}$.
 Taking $M=2$, we have
\begin{equation}
\label{eq:den}
 \delta = \frac{|g| f^4}{2\Lambda^4} \left(
\frac{f}{\sqrt{2}M_{Pl}} \right)^N \simeq \frac{f^4}{2
\Lambda^4}\left( \frac{f}{\sqrt{2}M_P} \right)^N,
\end{equation}
where we have taken $\left| g \right| \sim 1$.

For most of the studies in this paper we will assume that the
gravitational interaction correction to the mass of the Goldstone
scalar $\phi$ is less than that from the non-perturbative effects
{\it i.e.} $ \delta N^2  < 1$. This implies that $N$ in general is
expected to be large. For instance taking $f=0.5 M_{pl}$ and
$\Lambda =2.6 \times 10^{-4}M_{Pl} $, $ \delta N^2 < 1$ requires
that $N
> 36$, and for  $f=0.95 M_{pl}, \Lambda
=9.7 \times 10^{-4}M_{Pl}$ , one needs $N
> 90$. This requirement makes the
basic feature of the original natural inflation phenomenologically
unchanged, however theoretically it requires an convincing
argument to understand why the operators with dimensions less than
$N$ are forbidden or much suppressed. In this paper we will focus
on the phenomenological studies.

The inflaton potential under investigation in this paper is the
sum of $V_{planck}$ and the one in Eq.(\ref{eq:potentiala}),
namely
\begin{equation}
\label{eq:main}
 V(\phi)=\Lambda^4[1 + \cos(\phi/f)+ \delta \cos(N \phi /f +
\theta )].
\end{equation}
In additional to the two parameters $\Lambda$ and $f$ in the
original natural inflation we introduce in our model three more
parameters $\delta, N$ and $\theta$. The $\delta$ and $N$, however
are related by Eqs.$(5)$ and $(6)$.
And for simplicity of the discussions we will restrict ourself to
two cases that $\theta = 0$ or $\theta = \pi$ in this paper.
Choosing $\theta = \pi$ is equivalent to changing $\delta$ to
$-\delta$. Before concluding this section, we point out that
$\delta$ term shifts the minimum of the potential away from zero,
however since $\delta$ is
 extremely small, it will not change the results when minimizing the potential to zero,
  which we have made a numerical check for the calculation of the primordial spectrum and CMB results.

\section{Primordial Power Spectrum Calculation}

During inflation, the contribution of other matter can be neglected, and
the background evolution of the Universe is described by
\begin{equation}
\label{eq:H2}
 H^2=\frac{8\pi}{3M^2_{Pl}}\left[
V(\phi)+\frac{1}{2}\dot{\phi}^2 \right],
\end{equation}
\begin{equation}
\label{eq:accer}
 \ddot{\phi}+3H\dot{\phi}+V_{\phi} =0,
\end{equation}
where $H\equiv \dot{a}/a$ is the expansion rate, and $V_{\phi}
\equiv \partial V / \partial \phi$. Slow rolling (SR) requires
\begin{equation}
\label{eq:SRAcond1}
\epsilon \equiv
\frac{-\dot{H}}{H^2} \approx \frac{M_{Pl}^2}{16
\pi}(\frac{V_{\phi}}{V})^2 \ll 1,
\end{equation}
\begin{equation}
\label{eq:SRAcond2} \beta \equiv \frac{\ddot{\phi}}{H \dot{\phi}}
\approx \frac{M_{Pl}^2}{16 \pi}(\frac{V_{\phi}}{V})^2 -
\frac{M_{Pl}^2}{8 \pi}\frac{V_{\phi\phi}}{V} \ll 1.
\end{equation}
For the model under consideration in this paper, we have
\begin{equation}
\label{eq:SRA1}
 \epsilon \approx \frac{M_{Pl}^2}{16 \pi f^2 }
\frac{(\sin(\phi/f)+ \delta N \sin(N \phi/f))^2
}{\left(1+\cos(\phi/f)\right)^2},
\end{equation}

\begin{equation}
\label{eq:SRA2}
 \beta \approx \frac{M_{Pl}^2}{16\pi
f^2}\left(1+\frac{2\delta N^2 \cos (N \phi/f
)}{1+\cos(\phi/f)}\right).
\end{equation}

The primordial spectrum of the scalar perturbation is defined as
\begin{equation}
P_{S}(k) \equiv \frac{k^{3}}{2\pi^{2}} \left| \zeta_k
\right|^2,
\end{equation}
where $\zeta_k$ is the coefficient of the Fourier transform of the
Bardeen parameter $\zeta$ \cite{MFB92}. The scalar spectrum index
is given by
\begin{equation}
n_s(k)\equiv 1+\frac{d\ln P_s(k)}{d\ln k}.
\end{equation}

The primordial spectrum of the tensor perturbation is
\begin{equation}
P_{T}(k) \equiv \frac{k^{3}}{2\pi^{2}} \left| \psi_k \right|^2,
\end{equation}
where $\psi_k$ is the coefficient of the Fourier transform of
tensor linear perturbations $\psi$ \cite{MFB92}. For the tensor
spectrum index:
\begin{equation}
n_T(k)\equiv \frac{d\ln P_T(k)}{d\ln k}.
\end{equation}

In SR regime,
\begin{equation}
\label{eq:SRAspec} n_s=1-4\epsilon-2\beta, \qquad n_T=-2\epsilon.
\end{equation}
One can see from Eq.~(\ref{eq:SRA1}) and Eq.~(\ref{eq:SRA2}) that
the relative contributions of the $\delta$ term to the SR
parameter $\epsilon$ given by the terms of $\delta N$ or $\delta^2
N^2$ are proportional to $1/N$ or $\delta$ for a fixed $\delta
N^2$. Numerically they are small with the parameters (large $N$
and small $\delta$) we consider in this paper. But $\beta$ can be
strongly modulated by large $\delta
 N^2$, then $P_s(k)$ and $n_s$ are changed correspondingly. Note here
 $V$ and $V_\phi$ are changed minorly by $\delta$ term, hence the background evolution of the
 inflaton field will not be affected much.

 During inflation, the equation of motion for a Fourier mode of
fluctuation is \cite{MFB92}
\begin{equation}
\label{eq:modeODE}
u_{k}''+\left(k^{2}-\frac{z''}{z}\right)u_{k}=0,
\end{equation}
where
\begin{equation}
u \equiv z \zeta, \qquad z\equiv a\dot{\phi}/H.
\end{equation}
Here the prime denotes the derivative with respect to the
conformal time $d \eta \equiv dt/a(t)$. Well inside the horizon,
according to flat spacetime quantum field theory, the vacuum modes
are
\begin{equation}\label{eq:modeBCshort}
u_{k} \rightarrow
\frac{1}{\sqrt{2k}}e^{-ik\eta} ~~~\text{as}~~~~ k^2 \gg
\frac{z''}{z};
\end{equation}
Coming out of the horizon, the growing mode solution is
\begin{equation}
\label{eq:modeBClong} u_{k} \propto  z
~~~~~~~~~~~~~~~\text{for}~~~~ k^2 \ll \frac{z''}{z},
\end{equation}
with no explicit dependence on the behavior of scale factor $a$ in
this limit (frozen).
 For the slow roll approximation (SRA), $\frac{z''}{z}$ can
be expressed in terms of two SR parameters $\epsilon$ and $\beta$:
 \be
\frac{z''}{z}=2a^2H^2(1+\frac{3}{2}\beta+\epsilon+\frac{1}{2}\beta^2+
\frac{1}{2}\epsilon\beta+\frac{1}{2}\frac{1}{H}\dot{\epsilon}+
\frac{1}{2}\frac{1}{H}\dot{\beta}).
 \ee
If $\epsilon$ and $\beta$ can be approximatively considered as
constants, Eq.~(\ref{eq:modeODE}) can be rewritten as
\be
u_{k}''+\left(k^{2}-\frac{\nu^2-\frac{1}{4}}{\eta^2}\right)u_{k}=0,\lab{usr}
\ee
where $\nu=\frac{1+\beta+\epsilon}{1-\epsilon} + \frac{1}{2}$
and $\eta=\frac{-1}{aH}\left( \frac{1}{1-\epsilon}\right)$. The
solution of Eq.~(\ref{usr}) is a Bessel function,
 \be
u_k=\frac{1}{2}\sqrt{\pi}e^{i(\nu
+\frac{1}{2})\frac{\pi}{2}}(-\eta)^{\frac{1}{2}}H_{\nu}^{(1)}(-k\eta).\lab{ueom}
\ee
When $k/aH \rightarrow 0$, from the asymptotic form of
$H_{\nu}$ one obtains an expression of the scalar spectrum $P_S$:
\be
P_S(k,a) \rightarrow \frac{2^{2\nu-3}}{4 \pi^2
}\frac{\Gamma^2(\nu)}{\Gamma^2(\frac{3}{2})}(1-\epsilon)^{2\nu -1
}\frac{H^4}{\dot{\phi}^2}(\frac{k}{aH})^{3-2\nu},\lab{usr2}
\ee
where $P_S(k,a)$ varies with time. Traditionally one takes $k=aH$,
and gets the primordial spectrum \cite{SL}
 \be
\label{eq:usr3}
 P_S(k) \approx
\frac{2^{2\nu-3}}{4 \pi^2
}\frac{\Gamma^2(\nu)}{\Gamma^2(\frac{3}{2})}(1-\epsilon)^{2\nu -1
}\frac{H^4}{\dot{\phi}^2} \mid_{k=aH}.
\ee

The calculation for the gravitational wave spectrum is very
similar \cite{MFB92}. The equation of motion for tensor linear
perturbations is
\begin{equation}
\label{eq:T} v_k'' + (k^2 - \frac{a''}{a})v_k=0,
\end{equation}
where $v \equiv a \psi$, and
\begin{equation}
\label{eq:modeBCshortg} v_{k} \rightarrow
\frac{1}{\sqrt{2k}}e^{-ik\eta} ~~~\text{as}~~~~ k^2 \gg
\frac{a''}{a};
\end{equation}
\begin{equation}
\label{eq:modeBClongg} v_{k} \propto a
~~~~~~~~~~~~~~~\text{for}~~~ k^2 \ll \frac{a''}{a}.
\end{equation}
Therefore,
\begin{equation}
P_T(k)=\frac{k^3}{2 \pi^2}|\frac{v_k}{a}|^2.
\end{equation}
For SRA, one can get
\begin{equation}
\label{eq:PT} P_T(k)\approx 2^{2\mu-3}
\frac{\Gamma^2(\mu)}{\Gamma^2(\frac{3}{2})}(1-\epsilon)^{2\mu -1
}\frac{H^2}{4\pi^2} \mid_{k=aH},
\end{equation}
where
\be
\mu=\frac{1}{1-\epsilon}+\frac{1}{2}.
\ee

The Stewart-Lyth formula given in Eq.(\ref{eq:usr3}) and
Eq.(\ref{eq:PT}) are valid under certain conditions that both of
the parameters $\epsilon$ and $\beta$ are almost constants and the
asymptotic value of $P_S(k)$ can be approximated by $P_S(k=a H)$.
In Ref.\cite{WenBin} the authors pointed out a subtlety in the
approximations in obtaining the formula in Eq.(\ref{eq:usr3}),
however numerically they have also demonstrated that the
Stewart-Lyth formula works fairly well for the chaotic and natural
inflation models. In the presence of the $\delta$ term, the SRA is
not necessarily true and $\beta$ is oscillating, especially in the
case when $\delta N^2$ is relatively large. Thus, to ensure the
validity of Eq.(\ref{eq:SRAspec}), we calculate the spectrum
numerically in this paper.

  We numerically solve
 Eq.~(\ref{eq:H2}), Eq.~(\ref{eq:accer}), Eq.~(\ref{eq:modeODE})  and Eq.~(\ref{eq:T}) and obtain
the horizon crossing amplitude for each mode with the following
initial conditions \cite{WenBin,ACE01} :
\begin{eqnarray}
u_k(0) &=& \frac{1}{\sqrt{2k}}~ ,\\
u_k'(0) &=& \frac{-i k}{\sqrt{2k}}~ ;
\end{eqnarray}

\begin{eqnarray}
v_k(0) &=& \frac{1}{\sqrt{2k}}~ ,\\
v_k'(0) &=& \frac{-i k}{\sqrt{2k}}~ .
\end{eqnarray}

\begin{figure}[htbp]
\begin{center}
\includegraphics[scale=0.6]{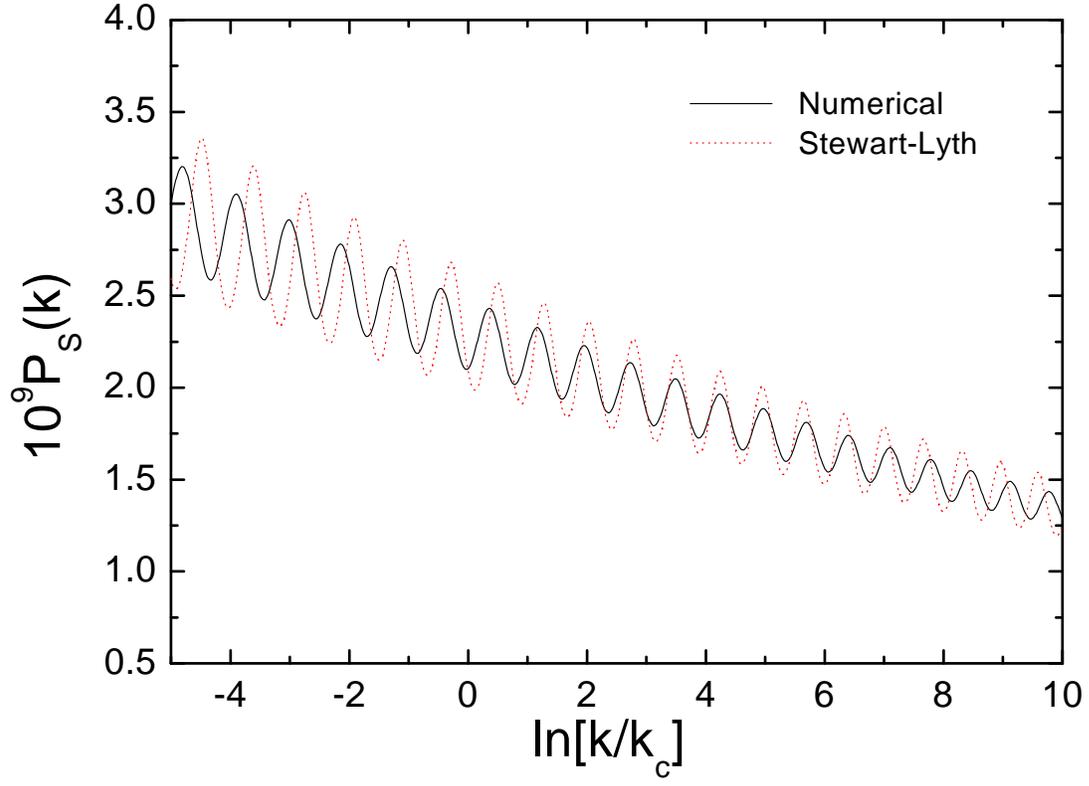}
\caption{\label{fig:contrast3} Plot of the power spectrum with
numerical results and Stewart-Lyth approximation. In the
calculation, we take $f=0.95 M_{pl}$, $\delta N^2 \approx 11$
($\delta=-3\times 10^{-5}, N=599$). $k_c$ is taken to be $k_c =
7.0a_0H_0$.}
\end{center}
\end{figure}

Our results show that for $\delta N^2 <1 $, the Stewart-Lyth
formula works well and the relative error is within $ 1\% $ in
comparison with the numerical results. However, for $\delta
N^2>1$, which we will discuss in detail in section V, the error
could be large. In Fig.\ref{fig:contrast3}, we plot $P_S(k)$ with
numerical results and the Stewart-Lyth approximation. One can see
that the difference is about 30 percent; for gravitational waves,
the difference is little.

\begin{figure}[htbp]
\begin{center}
\includegraphics[scale=0.35]{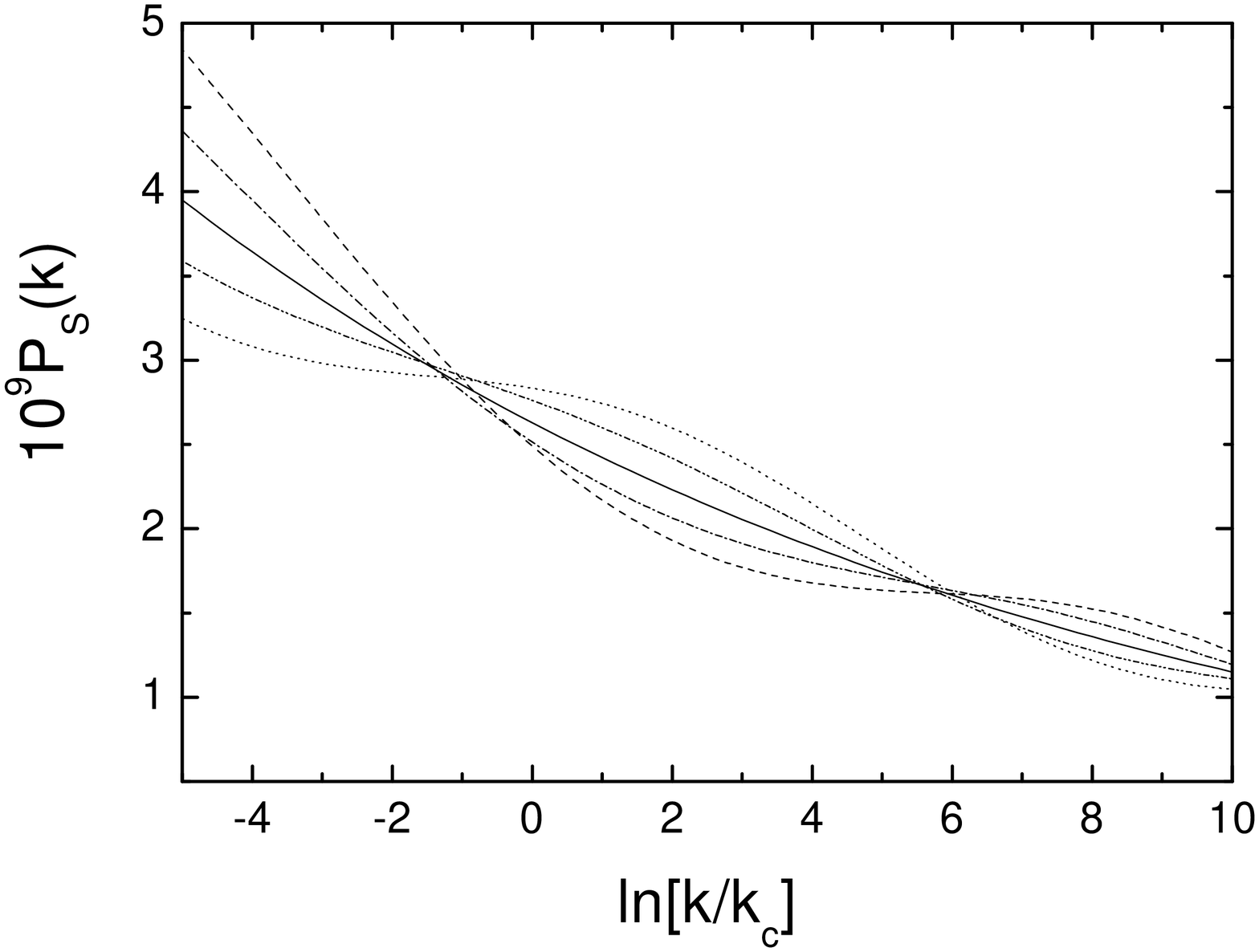}
\includegraphics[scale=0.35]{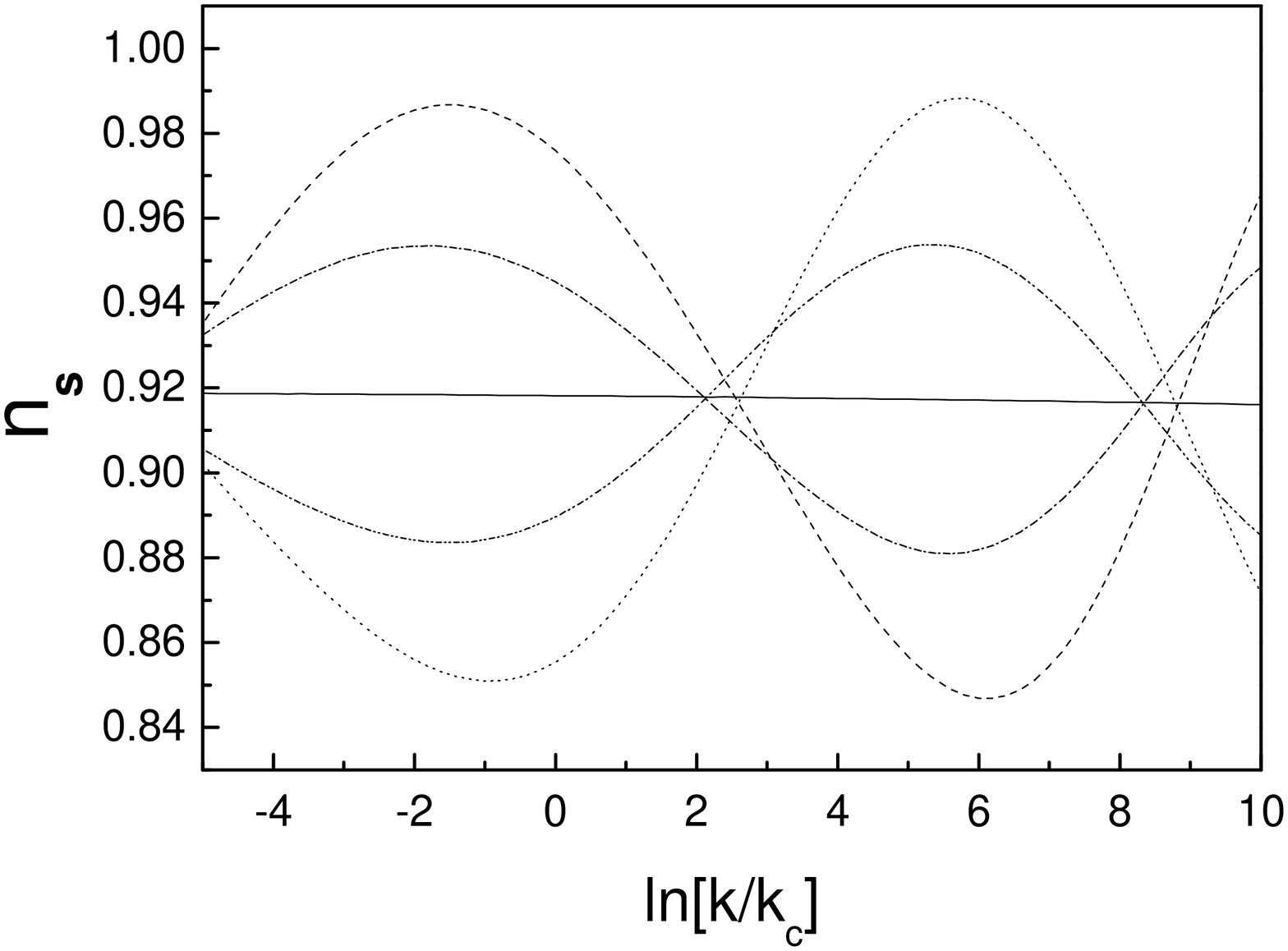}
 \caption{\label{fig:f7ns} Modulation of the power
spectrum and index, with $f=0.7M_{Pl}$, $\Lambda^4=2.0\times
10^{-13} M_{Pl}^4$. From left top to bottom, the lines stand for
$\delta N^2 =0.8, 0.42, 0, -0.42, -0.8$ respectively.}
\end{center}
\end{figure}

 In the natural inflation model, the scalar spectrum
index $n_s$ is generally smaller than 1. The $\delta$ term in our
inflaton potential induces a modulation on the power spectrum.
In Fig.~\ref{fig:f7ns}, we plot the primordial scalar spectrum and
index for a typical set of model parameters: $f=0.7M_{Pl}$, and
$\Lambda^4=2.0\times 10^{-13} M_{Pl}^4$. As can be seen from the
figures, the amplitude of the modulation in this model could be
$\mathcal{O}(10 \%)$.

 Before concluding this section we should point out
that the size of the modulation depends on the parameters of the
model, and $P_S(k)$ behaves quite differently for small and large
$f$. This
 can be understood qualitatively from the following two equations for natural inflation
 in the SR regime (Here we've set $8 \pi G =1$) :
 \be
 \label{eq:phi}
 \frac{d\phi}{d \ln a} \sim \frac{  \frac{1}{f}\sin(\phi/f)}
 {1+\cos(\phi/f)},
 \ee
 and
 \be
 \phi_{\mathcal{N}}=2f \arcsin(\frac{\sqrt{2}f}{\sqrt{1+2f^{2}}}  e^{
 -\frac{\mathcal{N}}{2f^{2}}}),
 \ee
where $\mathcal{N}$ is the number of $e$-$folding$ between  the
corresponding horizon exit and the end of inflation .
 For smaller $f$, $ \phi_{\mathcal{N}}$ would be much smaller than
 for larger $f$. For example $ \phi_{\mathcal{N}} \approx 2.03 $  for $f=0.95 M_{Pl}$,
 while $f=0.4 M_{Pl} $ gives $ \phi_{\mathcal{N}} \approx 6.3 \times 10^{-4} $
 for a given number of $e$-$folding$ $\mathcal{N}=70$. Such a small $\phi$ for $f=0.4 M_{Pl} $ would
also make $\frac{d\phi}{d \ln a}$ much smaller than $f=0.95 M_{Pl}
$. This is also true for our model when SRA is well satisfied.

\begin{figure}[htbp]
\begin{center}
\includegraphics[scale=0.35]{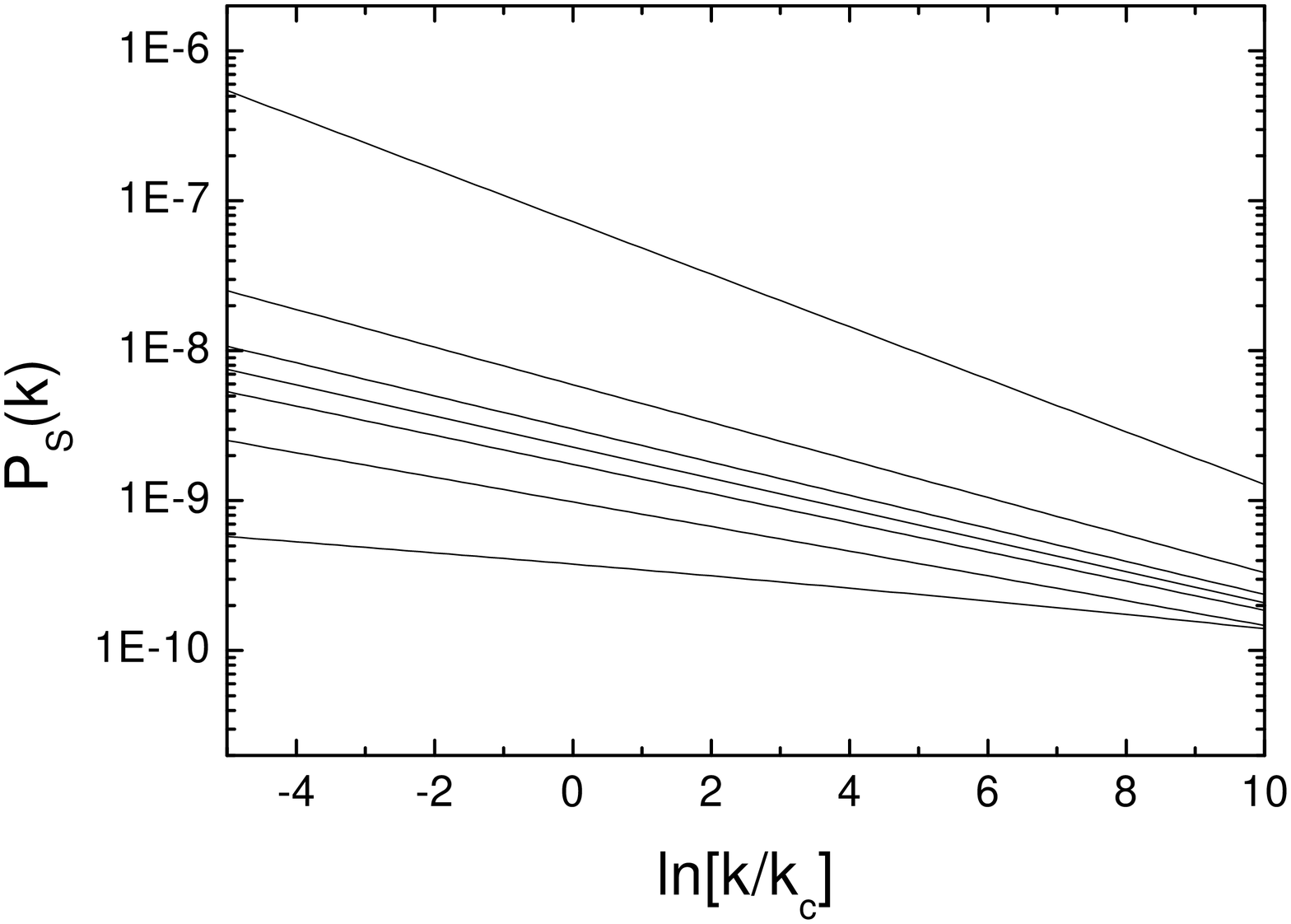}
\includegraphics[scale=0.35]{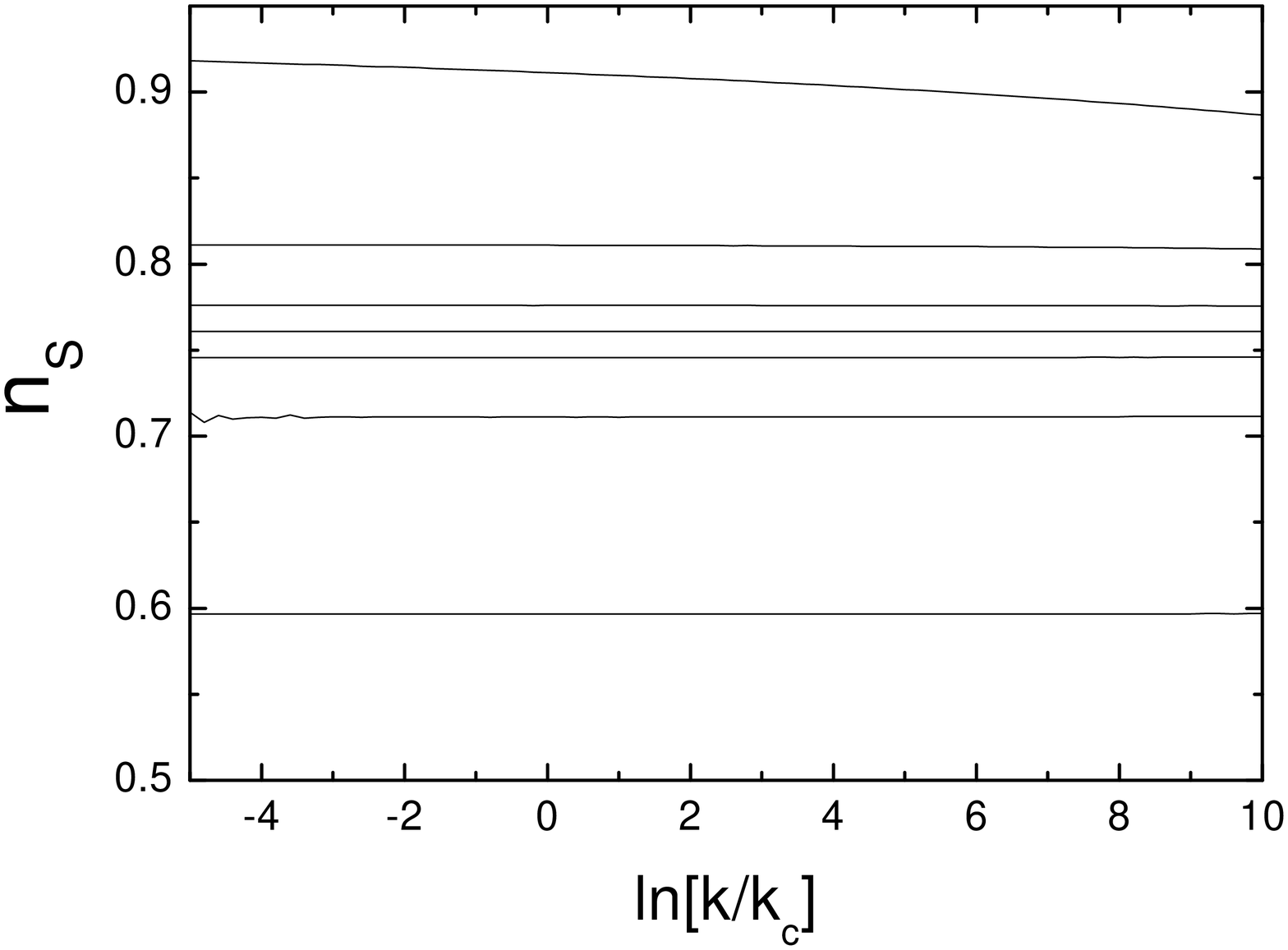}
 \caption{\label{fig:test4ns} The same as
Fig.\ref{fig:f7ns}, however with different model parameters,
$f=0.4 M_{Pl}$, $\Lambda^4=5.0\times 10^{-17} M_{Pl}^4$. From
above, the lines represent $\delta N^2$=$-0.7$, $-0.2$, $-0.07$,
$0$, $0.07$, $0.2$, $0.7$ respectively.}
\end{center}
\end{figure}

In Fig.\ref{fig:test4ns}, we plot the $P_S(k)$ and $n_S(k)$ for a
small $f$. One can see from this figure that for $f=0.4 M_{Pl}$,
$P_S(k)$ has lost the feature of oscillation in the range of scale
relevant to CMB and LSS, but the amplitudes of $P_S$ and $n_S$
change a lot.

\section{Results and comparison with observations}

\begin{figure}[htbp]
\begin{center}
\includegraphics[scale=0.35]{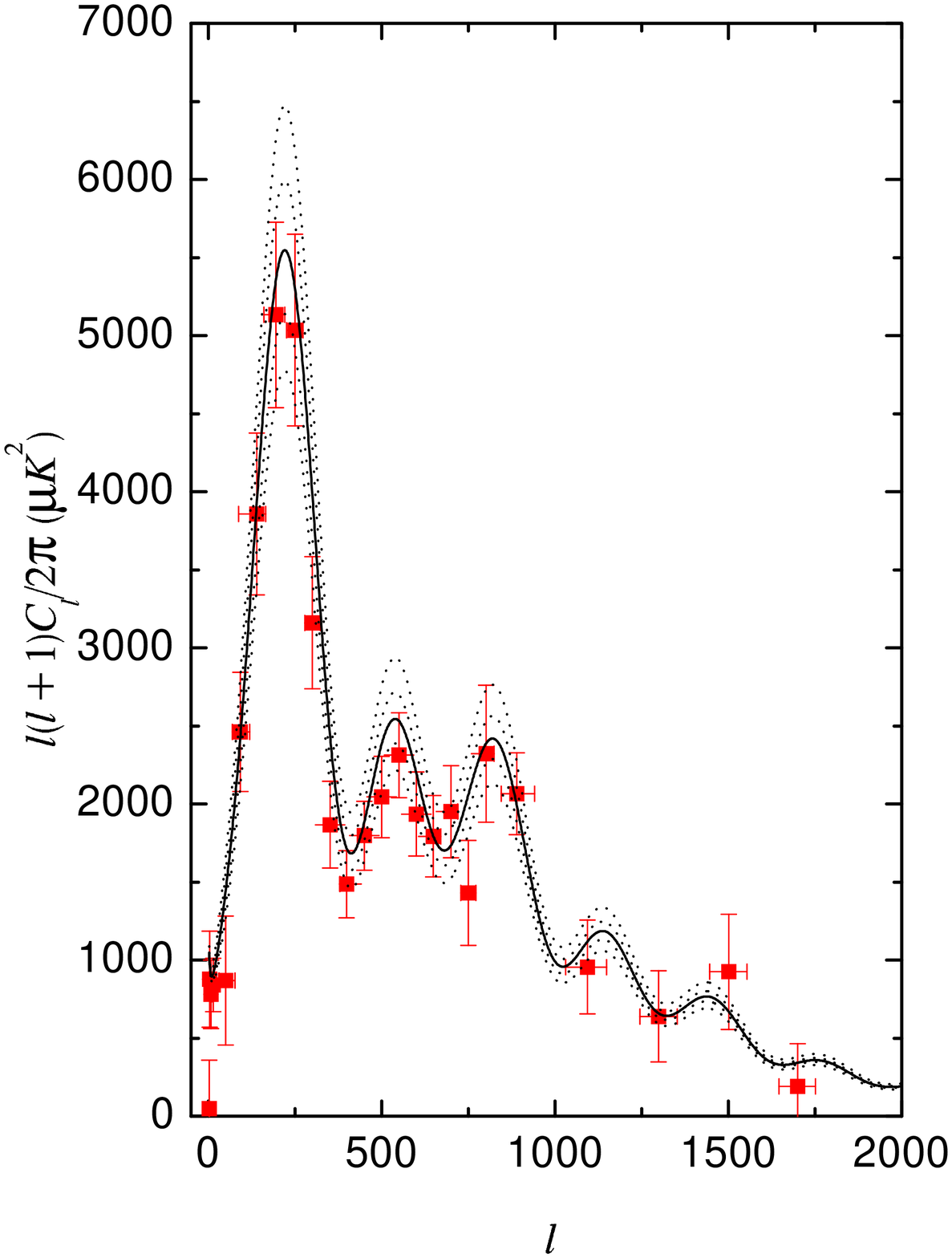}
\includegraphics[scale=0.35]{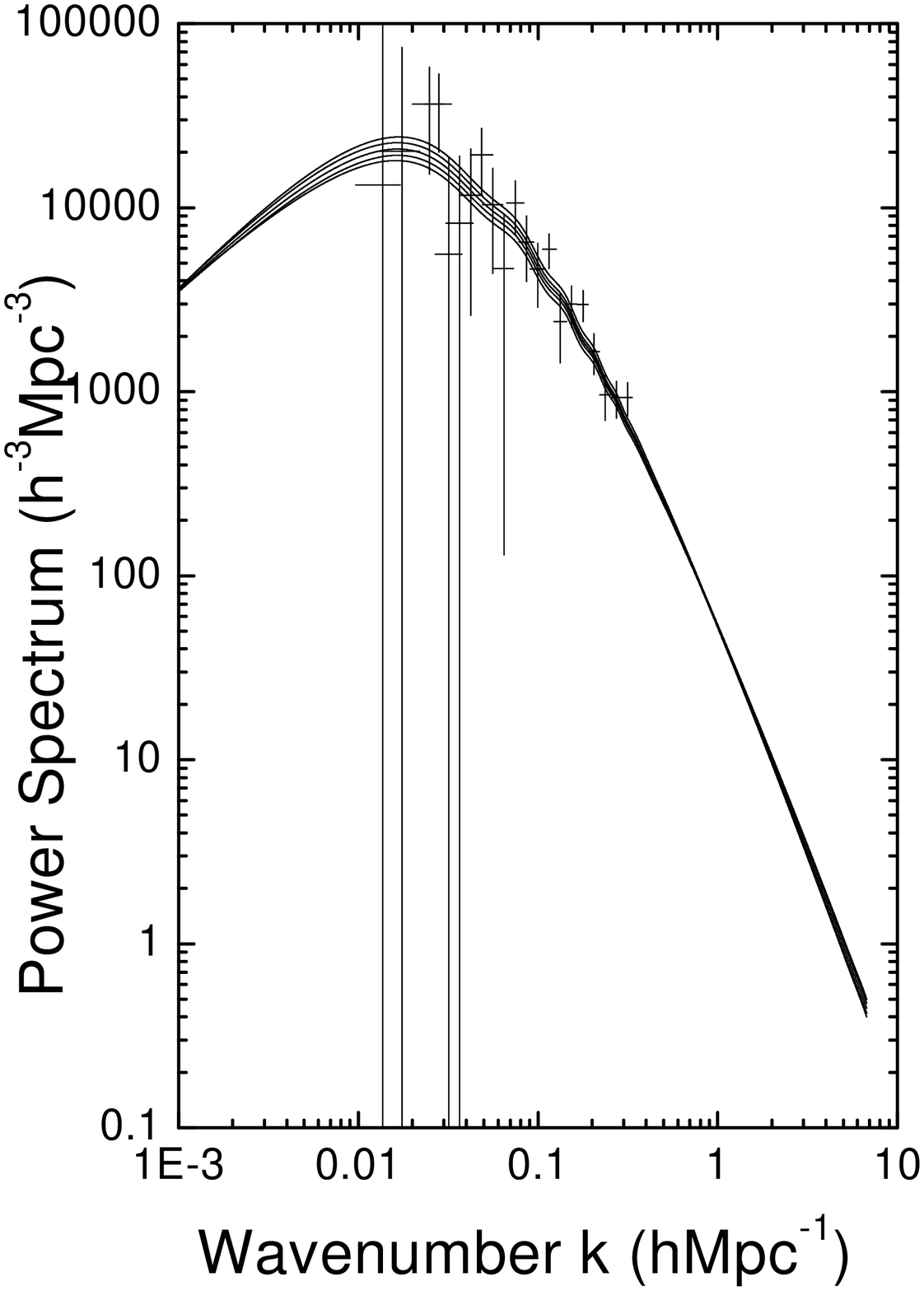}
 \caption{\label{fig:cmb750} The CMB anisotropy and
matter power spectrum for the parameters shown in
Fig.~\ref{fig:f7ns}, with $\delta N^2=0.8, 0.42, 0, -0.42, -0.8$
from above. The observational LSS data is from the PSCZ
catalogue\cite{HT}.}
\end{center}
\end{figure}

In this section we study the implications of our model on CMB and
LSS. The matter power spectrum is related to the primordial scalar
power spectrum by
\begin{equation}
\label{eq:LSS}
 P(k)=T^2 (k) P_S(k),
\end{equation}
where $T(k)$ is the matter transfer function, which can be
calculated for a given set of the cosmological background
parameters. The matter power spectrum in a large range of scales
can now be obtained by combining different data sets of CMB and
LSS measurements\cite{TZ02}. For example, currently the Lyman
alpha forest probes comoving scale of $0.1 \sim 10 hMpc^{-1}$, the
2dF galaxy correlation $0.01 \sim 0.1 hMpc^{-1}$, and CMB
measurements $0.001 \sim 0.1 hMpc^{-1}$. The CMB anisotropy is
related to the primordial scalar power spectrum by\cite{MB95}
\begin{equation}
\langle \Delta(\vec{n}_1) \Delta(\vec{n}_2)\rangle \equiv
\frac{1}{4\pi} \sum
\limits_{l=0}^{\infty}(2l+1)C_lP_l[\cos(\vec{n}_1 \cdot
\vec{n}_2)],
\end{equation}
\begin{equation}
\label {Cl}
 C_l\equiv \frac{2\pi}{l(l+1)}\tilde{C_l}=
\frac{4\pi}{(2l+1)} \int \frac{dk}{k}T_l^2(k)P_S(k),
\end{equation}
where $T_{l}(k)$ is the photon transfer function, which can be
calculated for a given set of the cosmological parameters. Here
$C_l$, $\tilde{C_l}$ and $T_l$ stand for the temperature angular
power spectrum and transfer function. We've omitted the subscript
$T$ for simplicity. We calculate $C_l$ by modifying the publicly
available {\tt CMBFAST} \cite{Seljak,CMBFAST} code. Similar
calculations for running $n_S$ have been performed in, for
example, Refs. \cite{Chung00,MT01,Covi}. The fiducial model
adopted in our calculation is the best fit model of
Ref.~\cite{Wang01}: $h=0.64, \Omega_{\Lambda}=0.66, \Omega_b
h^2=0.02$ and $\Omega_{k}=0$. We used the combined CMB data set
from Ref.\cite{MM02} in our plots. In Fig.~\ref{fig:cmb750} we
plot the CMB angular power spectrum and the matter power spectrum
with the primordial power spectrum shown in Fig.\ref{fig:f7ns}.
One can see that the theoretical predictions of our model are
quite different from that with $\delta = 0$ and the CMB data
disfavor the case with $\delta N^2= 0.8$.

\begin{figure}[htbp]

\newpage
\begin{center}
\includegraphics[scale=0.35]{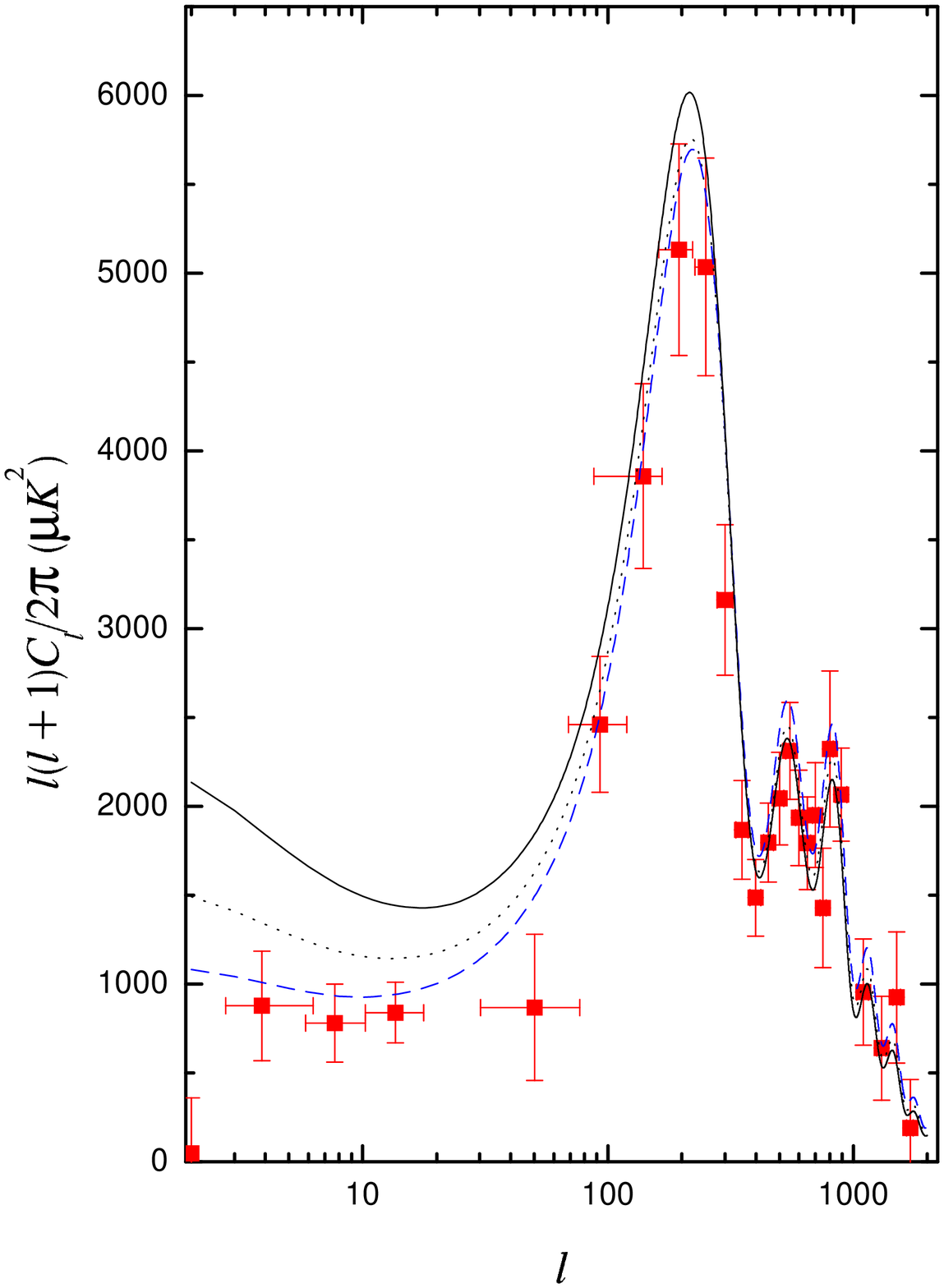}
\includegraphics[scale=0.35]{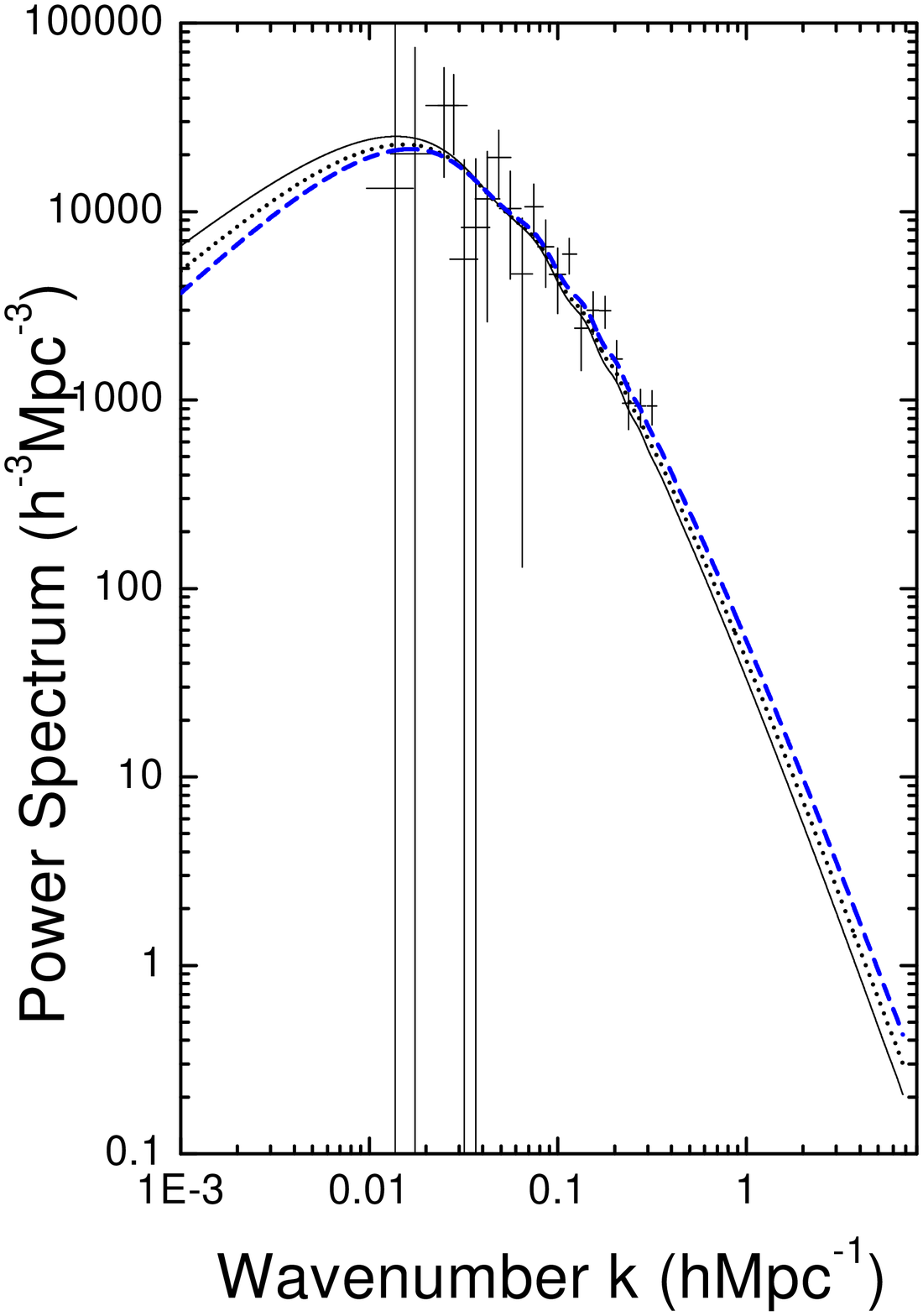}
\caption{\label{fig:f4cmbpk} $\tilde{C_l}$ (left) and matter power
spectrum (right) for a fixed $f=0.4 M_{Pl}$ but different values
of other model parameters. For the solid line,
$\Lambda^4=1.3\times 10^{-16} M_{Pl}^4$, $\delta N^2=0$; The
dotted line stands for $\Lambda^4=3.6\times 10^{-16} M_{Pl}^4$,
$\delta N^2=-0.34$; For the dashed line, $\Lambda^4=5.2\times
10^{-16} M_{Pl}^4$, $\delta N^2=-0.78$. }
\end{center}
\end{figure}

One interesting point of our model is the running behavior of
$n_s$ in the case with a small $f$, which we have shown in
Fig.~\ref{fig:test4ns}. One can see that without $\delta$ term the
model predicts $n_s$ below 0.8, which is ruled out already by the
data. However with the help of the $\delta$ term the theoretical
predictions can be made to be consistent with the observations
which can be seen from Fig.~\ref{fig:f4cmbpk}. Numerically in
natural inflation with $f=0.4M_{pl}$ it predicts $n_S\approx
0.76$. And the fine tuning of the $V_0$ would not be able to make
the theory consistent with $C_l$ and $P(k)$ \cite{Wang01}. However
the presence of the $\delta$ term makes the fit better.

Up to now we have restricted ourself to the parameter space
$\delta N^2 <1$, however it would be interesting to take the
potential in Eq.~(\ref{eq:main}) as a phenomenological model and
study the cosmological implications with an extended parameter
space $\delta N^2 > 1$ . In this case, there are four parameters,
where we take $f$ , $N$ and $\delta$ free, however $\Lambda$ can
be normalized by observations.

\begin{figure}[htbp]
\begin{center}
\includegraphics[scale=0.6]{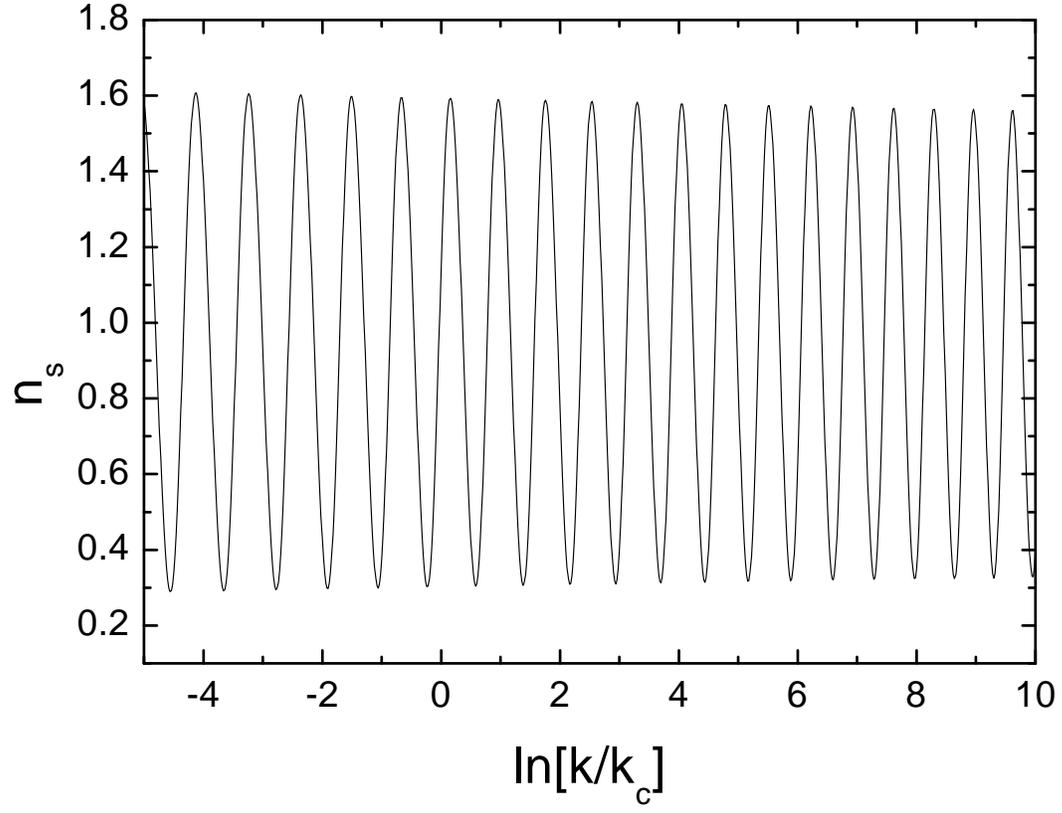}
 \caption{\label{fig:ns955} The primordial power
spectrum index for $f=0.95 M_{Pl}$, $N=599$, $\delta=-3 \times
10^{-5}$.}
\end{center}
\end{figure}

\begin{figure}[htbp]
\begin{center}
\includegraphics[scale=0.6]{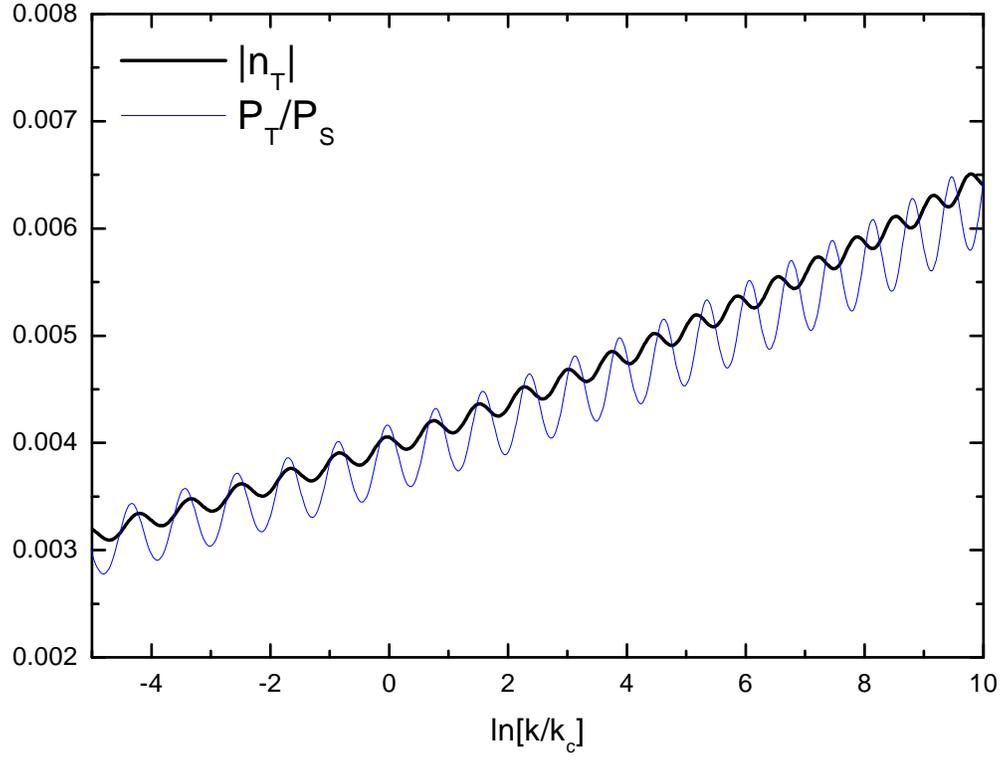}
 \caption{\label{fig:-9503nT} Plot of $|n_T|$ and $P_T/P_S$ for $f=0.95 M_{Pl}$, $N=599$ and $\delta=-3 \times
10^{-5}$.}
\end{center}
\end{figure}

\begin{figure}[htbp]
\begin{center}
\includegraphics[scale=0.6]{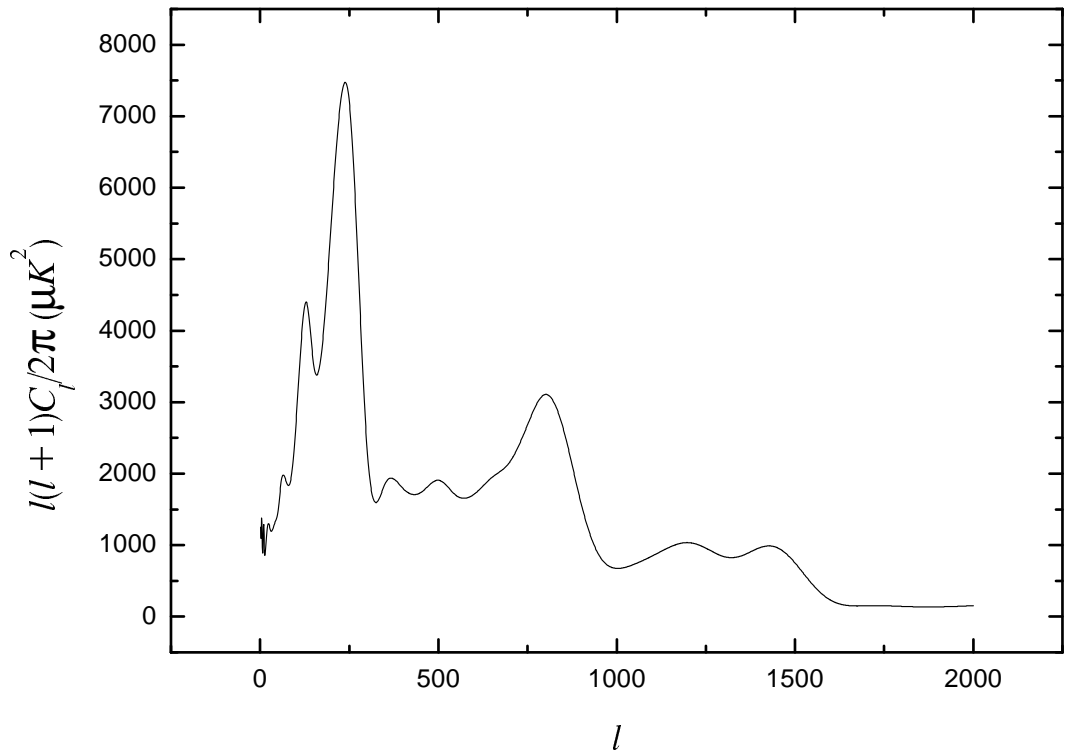}
\caption{\label{fig:2ndpeak} Theoretical prediction on $C_l$ with
$f=.95 M_{pl}$, $\delta=0.0007$  and $N=799$. }
\end{center}
\end{figure}

The effect of the modulation on the power spectrum depends on the
amplitude, frequency, and phase. And the effect  is more apparent
for a high frequency (large $N$), as can be seen in
Figs.~\ref{fig:ns955}-\ref{fig:polarization}. In
Fig.~\ref{fig:ns955} we take $N=599$ and find the variation of
primordial scalar spectrum index is very frequent and its
amplitude changes from about $0.3$ to $1.6$, although the value of
$\delta$ is very small. However, our results also show that the
tensor perturbation $P_T$ changes little since $\epsilon\ll 1 $.
We have checked the validity of the consistency
relation\cite{CR0}\footnote{The consistency relation holds for
single-field SRA inflation. The authors of Ref.\cite{CR} argue
that the trans-Planckian physics may change the vacuum and
provides an example for the violation of consistency relation.}.
One can see from Fig.~\ref{fig:-9503nT} for different range of
$k$, $|n_T|$ is larger or smaller than $P_T/P_S$. The reason
responsible for the violation of consistency relation in our model
is due to a large running of $\beta$ from $- 0.3$ to $0.3$ , which
indicates SRA isn't satisfied well. However, we will show below
that this type of primordial spectrum doesn't contradict to the
current observations. In Fig.\ref{fig:tmp} we plot the CMB
anisotropy and the matter power spectrum for the given parameters
and find that there can be a clear modulation on the matter power
spectrum, which is somewhat similar to the baryonic oscillation
wiggles, but has an entirely different origin. Of course, at
present there is no observational evidence of such wiggles in the
power spectrum. The peaks found in recent CMB measurements seem to
agree reasonably well with the predictions of power law primordial
spectrum. In the large scale structure data, there is some
tentative report on wiggles in the power spectrum
\cite{P01,M01a,M01b,H02}. These are generally thought to be due to
baryonic oscillation, but the effect discussed here could also
give rise to such wiggles. Since the position and frequency of the
baryonic wiggles can be predicted, and there is no reason to
expect that the primordial power spectrum modulation to coincide
with it, there is hope to distinguish these two cases with high
precision data. However, we note that at present time there is
still no clear evidence even for the baryonic wiggles
\cite{Xu,MNC02}. Note that the modulation on the power spectrum is
coherent, with different $P_S$ the shape of the CMB peaks can be
also quite different, some would drastically change the structure
of the $\tilde{C_l}$ spectrum, the peaks of the $\tilde{C_l}$ can
be  split into two in some particular cases. In the left panel of
Fig.\ref{fig:tmp}, the first peaks have been clearly split, while
in Fig.\ref{fig:2ndpeak}, it shows that the original first and
second peaks have been split, the third peak is highly raised and
new peaks are also generated. In Fig.~\ref{fig:polarization} we
plot the CMB polarizations which are given by
\begin{equation}
\label {ClE}
 C_{El}=\frac{4\pi}{(2l+1)} \int \frac{dk}{k}T_{El}^2(k)P_S(k),
\end{equation}
\begin{equation}
\label {CCl}
 C_{Cl}=\frac{4\pi}{(2l+1)} \int
 \frac{dk}{k}T_{l}(k)T_{El}(k)P_S(k)
\end{equation}
for the E-mode and cross correlation spectrum polarization.
 These effects might be
detectable by further CMB observations\cite{Plank}.

\begin{figure}[htbp]
\begin{center}
\includegraphics[scale=0.6]{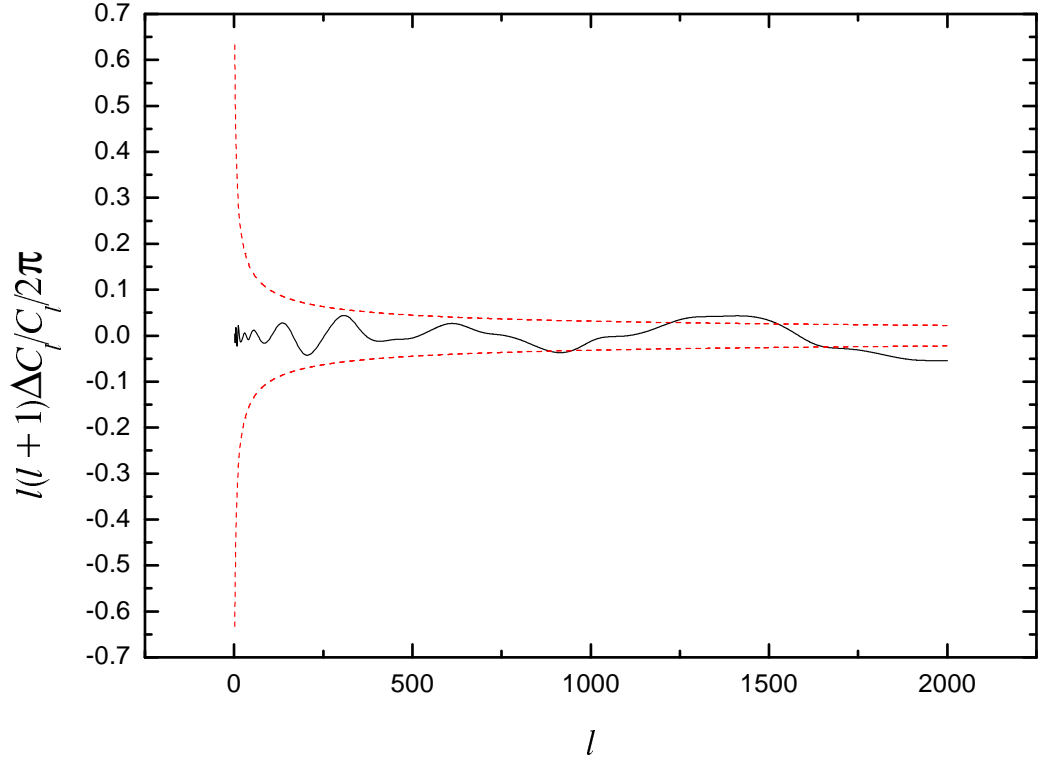}
\caption{\label{fig:diffcl} Plot of $\Delta \tilde{C_l}$  and the
cosmic variance limits. }
\end{center}
\end{figure}

Current CMB  observations do not exclude entirely the oscillating
primordial spectrums. Consider the difference between
$\tilde{C_l^0}$ and $\tilde{C}_l ^
 1$ where $\tilde{C_l^0}$ is calculated with $f=0.95 M_{pl}$,
 $\delta=0$ ($n_S\approx 0.945$) and $\tilde{C}_l ^1$ with $f=0.95 M_{pl}$ , $N=599$ ,
  $\delta=-3\times 10^{-5}$, then compare them with the cosmic
  variance\cite{variance} limits on $\tilde{C_l^0}$. In Fig.\ref{fig:diffcl},
  the solid line stands for $\frac{\tilde{C}_l ^ 1-\tilde{C}_l^0}{\tilde{C}_l^0}$
and the region  between the dashed lines is given by the cosmic
variance limits. Here $\tilde{C}_l ^ 1$ and $\tilde{C}_l^0$ are
normalized by COBE. Cosmic variance plays a fundamental limit to
the $\tilde{C_l}$ spectrum that can be measured, with

\begin{equation}
\label{eqn:CV}
 \frac{\Delta \tilde{C_l}}{\tilde{C_l}}  \geq \sqrt{ \frac{2}{2 l +1}} .
\end{equation}
One can see from Fig.\ref{fig:diffcl} that $\tilde{C}_l ^ 1$ and
$\tilde{C}_l^0$ are hardly distinguishable. Even though the
primordial spectrum (see, Fig.\ref{fig:ns955}) is oscillating, its
contribution to CMB, because of being averaged over multiple l,
becomes insignificant in this case.

\section{Discussions and Summary}

\begin{figure}[htbp]
\begin{center}
\includegraphics[scale=0.35]{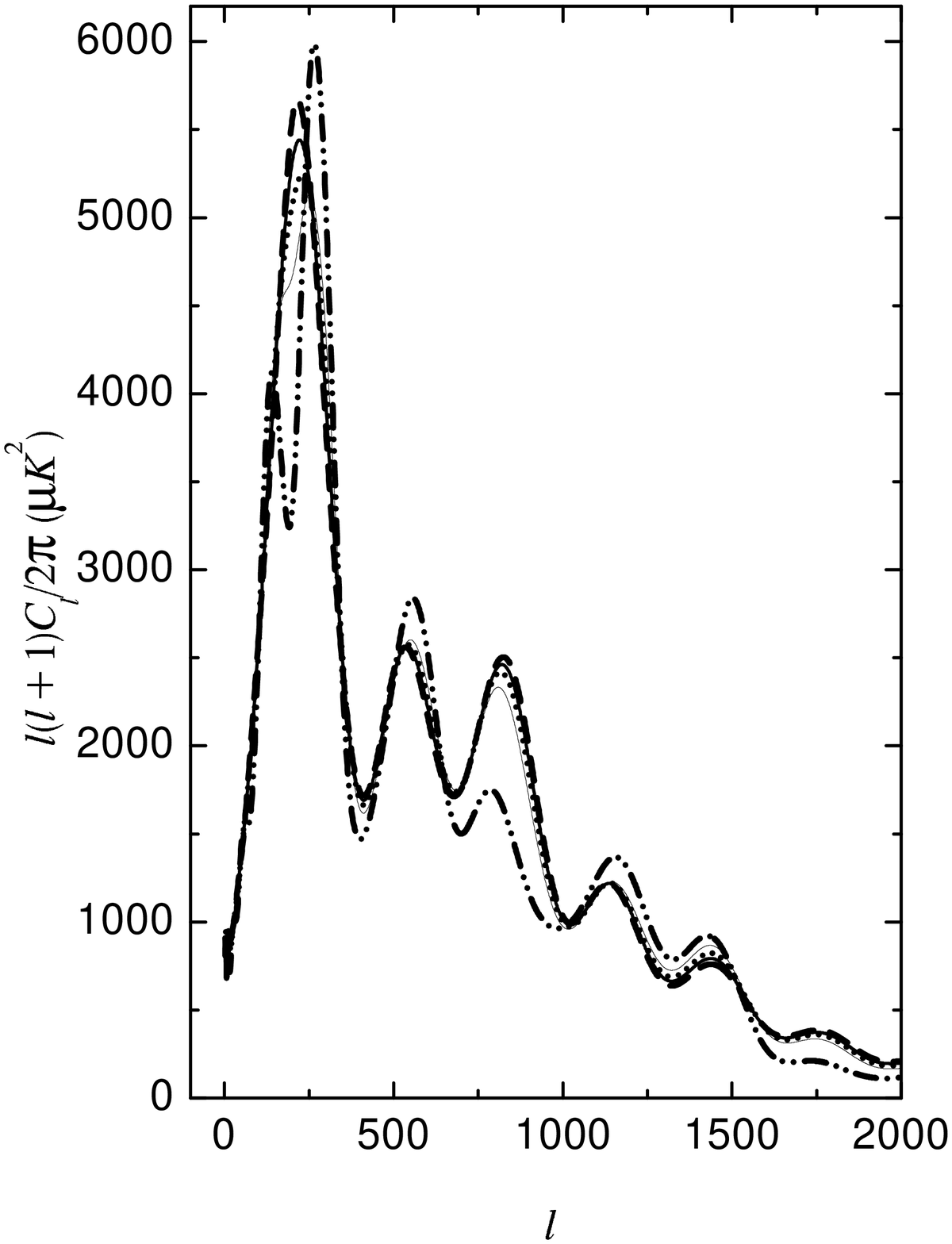}
\includegraphics[scale=0.35]{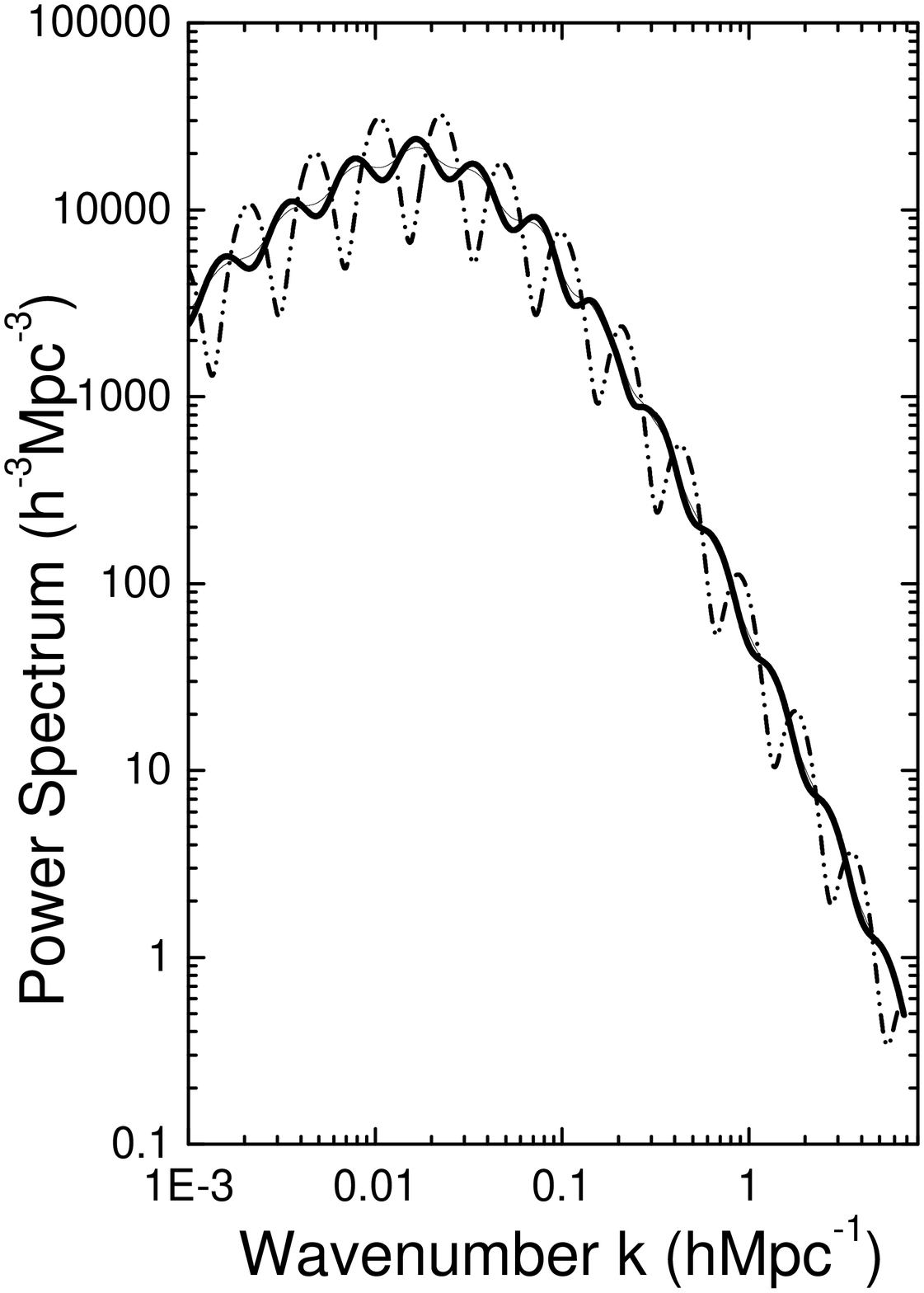}
 \caption{\label{fig:tmp} CMB anisotropy and matter
power spectrum for $f=0.95 M_{Pl}, N=599$. Left panel: the dash
dot dotted, the dashed, the dotted, the thicker solid and the
thinner solid line stand for $\delta=-3\times 10^{-4}$, $\delta=
3\times 10^{-5}$, $\delta= -3\times 10^{-5}$, $\delta=0$ and
$\delta=-8\times 10^{-5}$ respectively; Right panel: The dash
dotted dot, the thicker solid and the thinner solid stand for
$\delta=-3\times 10^{-4}$, $\delta= -8\times 10^{-5}$ and
$\delta=-3\times 10^{-5}$ respectively.
 }
\end{center}
\end{figure}

\begin{figure}[htbp]
\begin{center}
\includegraphics[scale=0.35]{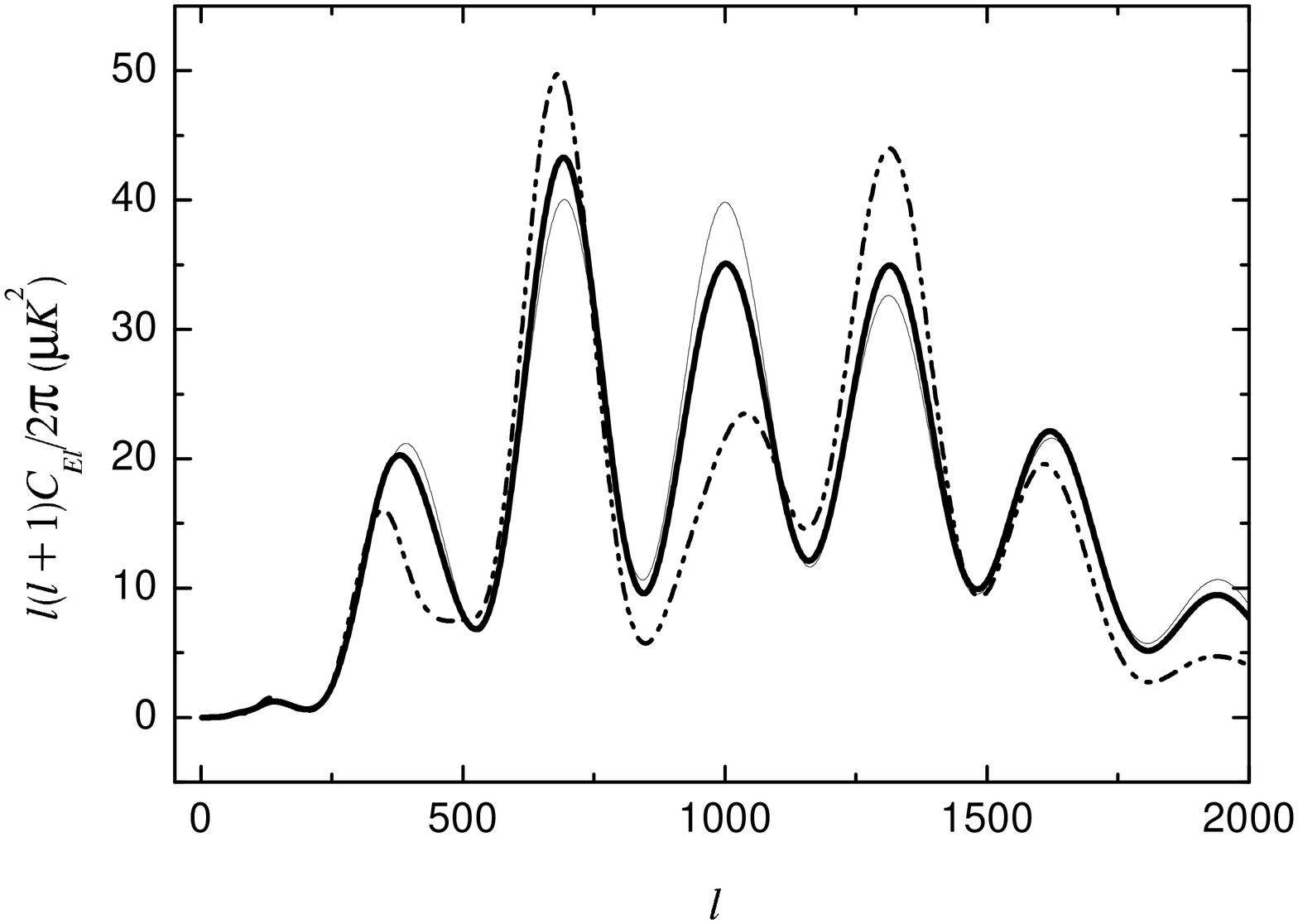}
\includegraphics[scale=0.35]{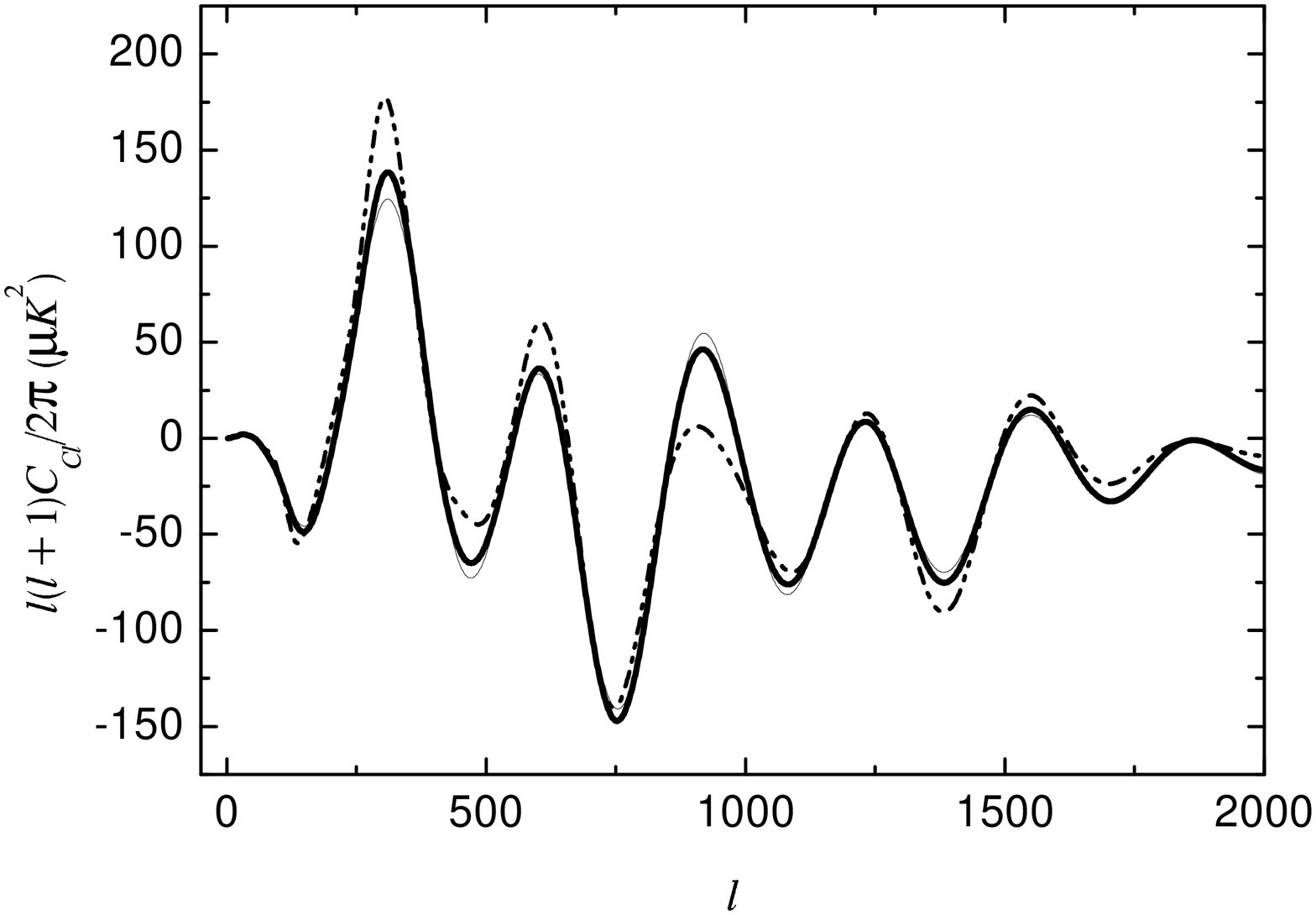}
 \caption{ \label{fig:polarization} CMB polarization for the same
parameters as in the right panel of Fig. \ref{fig:tmp} .}
\end{center}
\end{figure}

In this paper we present a  model which is a variation of natural
inflation. We have shown two features of the primordial spectrum
of this model, oscillating and scale-dependence and studied the
implications on CMB and LSS. In the presence of the $\delta$ term
the parameter space allowed for a successful natural inflation
will be enlarged relative to the original natural inflation
model\cite{MT01}. When the parameter space is enlarged and
extended to $\delta N^2 > 1$ for example due to some other
physical motivations\cite{FLZZ}, there are several additional
interesting effects. Although the SR approximation is violated and
the spectral index oscillates with a large scale variation, there
could be a large parameter space not ruled out by the
observations. As we can see from Fig.\ref{fig:tmp} when we
gradually increase the value of $\delta$ the effects on CMB
firstly take place on the first peak which can be slightly split,
meanwhile wiggles on the matter power spectrum are gradually
enhanced. The effects on CMB first peak and large scale structure
are potentially observable and can be tested by future precise
experiments.

In summary we have studied in this paper a model which is a
variation of natural inflation and show that there are some
interesting phenomenological features of this  model, such as
oscillating and scale-dependence in the primordial spectrums. And
we have also discussed their implications on CMB and LSS.

{\bf Acknowledgments} We thank  R. Brandenberger  for comments and
suggestions on the manuscript. This work is supported in part by
National Natural Science Foundation of China under Grant
No.90303004 and by Ministry of Science and Technology of China
under Grant No.NKBRSF G19990754.

\newcommand\AJ[3]{~Astron. J.{\bf ~#1}, #2~(#3)}
\newcommand\APJ[3]{~Astrophys. J.{\bf ~#1}, #2~ (#3)}
\newcommand\APJL[3]{~Astrophys. J. Lett. {\bf ~#1}, L#2~(#3)}
\newcommand\APP[3]{~Astropart. Phys. {\bf ~#1}, #2~(#3)}
\newcommand\CQG[3]{~Class. Quant. Grav.{\bf ~#1}, #2~(#3)}
\newcommand\JETPL[3]{~JETP. Lett.{\bf ~#1}, #2~(#3)}
\newcommand\MNRAS[3]{~Mon. Not. R. Astron. Soc.{\bf ~#1}, #2~(#3)}
\newcommand\MPLA[3]{~Mod. Phys. Lett. A{\bf ~#1}, #2~(#3)}
\newcommand\NAT[3]{~Nature{\bf ~#1}, #2~(#3)}
\newcommand\NPB[3]{~Nucl. Phys. B{\bf ~#1}, #2~(#3)}
\newcommand\PLB[3]{~Phys. Lett. B{\bf ~#1}, #2~(#3)}
\newcommand\PR[3]{~Phys. Rev.{\bf ~#1}, #2~(#3)}
\newcommand\PRL[3]{~Phys. Rev. Lett.{\bf ~#1}, #2~(#3)}
\newcommand\PRD[3]{~Phys. Rev. D{\bf ~#1}, #2~(#3)}
\newcommand\PROG[3]{~Prog. Theor. Phys.{\bf ~#1}, #2~(#3)}
\newcommand\PRPT[3]{~Phys.Rept.{\bf ~#1}, #2~(#3)}
\newcommand\RMP[3]{~Rev. Mod. Phys.{\bf ~#1}, #2~(#3)}
\newcommand\SCI[3]{~Science{\bf ~#1}, #2~(#3)}
\newcommand\SAL[3]{~Sov. Astron. Lett{\bf ~#1}, #2~(#3)}

\newpage

\newpage

\newpage

\newpage

\newpage

\newpage

\newpage

\newpage

\newpage

\newpage

\end{document}